\documentclass[longauth]{aa}

\usepackage{multirow}
\usepackage{graphicx}
\usepackage{txfonts}
\usepackage{longtable}
\usepackage{array}
\usepackage{amsmath}
\usepackage{hyperref}        

\usepackage{lscape}             
\usepackage{placeins}           
\linenumbers
\renewcommand\makeLineNumber{}                               

\begin{document}

   \title{Tomography analysis of the intermediate-luminosity Type Iax SN 2024pxl}

   \titlerunning{The type Iax SN 2024pxl}

   \author{
Barnabás Barna\inst{1,2} \and
Mridweeka Singh\inst{3} \and
Lindsey A.\ Kwok\inst{4} \thanks{CIERA Fellow} \and
Saurabh W.\ Jha\inst{5} \and
R.~Dastidar\inst{6} \and
Conor~Larison\inst{5} \and
Alexei V.\ Filippenko\inst{7} \and
J.~P. Anderson\inst{8} \and
Jennifer E.\ Andrews \inst{9} \and
Moira~Andrews\inst{10,11} \and
G. C. Anupama\inst{12} \and
Prasiddha Arunachalam\inst{13} \and
Katie Auchettl\inst{14,14} \and
Dominik Bánhidi\inst{1,15} \and
K.\ Azalee Bostroem\inst{16} \thanks{LSST-DA Catalyst Fellow} \and
Thomas G.\ Brink\inst{7} \and
Régis Cartier\inst{17} \and
Ping Chen\inst{18,19} \and
T.-W. Chen\inst{20} \and
Collin~T.~Christy\inst{16} \and
David~A.~Coulter\inst{21} \and
Sofia~Covarrubias\inst{22} \and
Kyle W.\ Davis\inst{13} \and
Connor~B.~Dickinson\inst{13} \and
Yize Dong\inst{23} \and
Joseph Farah\inst{10,11} \and
Andreas Flörs\inst{24} \and
Ryan J.\ Foley\inst{13} \and
Noah Franz\inst{16} \and
Christoffer~Fremling\inst{22,25} \and
Lluís Galbany\inst{26,27} \and
Anjasha Gangopadhyay\inst{28} \and
Aarna~Garg\inst{13} \and
Elinor~L.~Gates\inst{29} \and
Or Graur\inst{30,31} \and
Mariusz Gromadzki\inst{32} \and
Daichi~Hiramatsu\inst{23,33} \and
Emily~Hoang\inst{34} \and
D.~Andrew~Howell\inst{10,11} \and
Brian Hsu\inst{16} \and
Joel Johansson\inst{35} \and
Arti Joshi\inst{36} \and
Lordrick~A.~Kahinga\inst{13} \and
Ravjit Kaur\inst{13} \and
Sahana~Kumar\inst{37} \and
Piramon~Kumnurdmanee\inst{13} \and
Hanindyo Kuncarayakti\inst{38,39} \and
Chang~Liu\inst{4,40} \and
Keiichi Maeda\inst{41} \and
Kate Maguire\inst{42} \and
Bailey Martin\inst{43} \and
Curtis McCully\inst{10,11} \and
Darshana Mehta\inst{34} \and
Luca~M.~Menotti\inst{13} \and
Anne~J.~Metevier\inst{44} \and
A.~A.~Miller\inst{4,33,45} \and
Kuntal Misra\inst{46} \and
Erika Mochnács\inst{1} \and
T. E. Müller-Bravo\inst{42,47} \and
Megan Newsome\inst{10,48} \and
Estefania Padilla~Gonzalez\inst{49} \and
Kishore~C.~Patra\inst{13} \and
Jeniveve Pearson\inst{16} \and
Anthony~L.~Piro\inst{50} \and
Abigail~Polin\inst{51} \and
Aravind~P.~Ravi\inst{35} \and
Armin Rest\inst{21,49} \and
Nicolas~Meza~Retamal\inst{35} \and
O.~M.~Robinson\inst{13} \and
César Rojas-Bravo\inst{52,53} \and
Devendra~K.~Sahu\inst{12} \and
David~J.~Sand\inst{16} \and
Brian~P.~Schmidt\inst{43} \and
Steve Schulze\inst{4} \and
Michaela Schwab\inst{5} \and
Manisha Shrestha\inst{16} \and
Matthew~R.~Siebert\inst{21} \and
Sunil Simha\inst{4,54} \and
Nathan Smith\inst{16} \and
Jesper Sollerman\inst{28} \and
Shubham~Srivastav\inst{55} \and
Bhagya~M.~Subrayan\inst{16} \and
Tamás Szalai\inst{1,56} \and
Kirsty Taggart\inst{13} \and
Rishabh Singh Teja\inst{12} \and
Jacco~H.~Terwel\inst{42} \and
Samaporn Tinyanont\inst{57} \and
Stefano Valenti\inst{35} \and
József Vinkó\inst{1,48,58,59} \and
Aya L. Westerling\inst{13} \and
J.\ Craig Wheeler\inst{48} \and
Yi Yang\inst{7,60} \and
WeiKang Zheng\inst{7}
}

\institute{
Department of Experimental Physics, University of Szeged, D\'om t\'er 9, Szeged 6720, Hungary \and
HUN-REN-SZTE Stellar Astrophysics Research Group, 6500 Baja, Szegedi út, Kt. 766, Hungary \and
Indian Institute of Astrophysics, Koramangala 2nd Block, Bangalore 560034, India \and
Center for Interdisciplinary Exploration and Research in Astrophysics (CIERA), 1800 Sherman Ave., Evanston, IL 60201, USA \and
Department of Physics and Astronomy, Rutgers, the State University of New Jersey,\\136 Frelinghuysen Road, Piscataway, NJ 08854-8019, USA \and
Istituto Nazionale di Astrofisica, Osservatorio Astronomico di Brera, via E. Bianchi 46, 23807 Merate (LC), Italy \and
Department of Astronomy, University of California, Berkeley, CA 94720-3411, USA \and
European Southern Observatory, Alonso de Córdova 3107, Vitacura, Casilla 19001, Santiago, Chile \and
Gemini Observatory/NSF's NOIRLab, 670 North A`ohoku Place, Hilo, HI 96720-2700, USA \and
Las Cumbres Observatory, 6740 Cortona Drive, Suite 102, Goleta, CA 93117-5575, USA \and
Department of Physics, University of California, Santa Barbara, CA 93106-9530, USA \and
Indian Institute of Astrophysics, Koramangala 2nd Block, Bangalore 560034, India \and
Department of Astronomy and Astrophysics, University of California, Santa Cruz, CA 95064-1077, USA \and
School of Physics, The University of Melbourne, Parkville, VIC 3010, Australia \and
Baja Astronomical Observatory of the University of Szeged, Szegedi {\'u}t, Kt. 766, 6500 Baja, Hungary \and
Steward Observatory, University of Arizona, 933 North Cherry Avenue, Tucson, AZ 85721-0065, USA \and
Centro de Astronomía (CITEVA), Universidad de Antofagasta, Av. Angamos 601, Antofagasta, Chile \and
Institute for Advanced Study in Physics, Zhejiang University, Hangzhou 310027, China \and
Department of Particle Physics and Astrophysics, Weizmann Institute of Science, 76100 Rehovot, Israel \and
Graduate Institute of Astronomy, National Central University, 300 Jhongda Road, 32001 Jhongli, Taiwan \and
Space Telescope Science Institute, 3700 San Martin Drive, Baltimore, MD 21218-2410, USA \and
Division of Physics, Mathematics and Astronomy, California Institute of Technology, Pasadena, CA 91125, USA \and
Center for Astrophysics Harvard \& Smithsonian, 60 Garden Street, Cambridge, MA 02138-1516, USA \and
GSI Helmholtzzentrum f\"ur Schwerionenforschung, Planckstra\ss{}e 1, 64291 Darmstadt, Germany \and
Caltech Optical Observatories, California Institute of Technology, Pasadena, CA 91125, USA \and
Institute of Space Sciences (ICE, CSIC), Campus UAB, Carrer de Can Magrans, s/n, E-08193 Barcelona, Spain \and
Institut d’Estudis Espacials de Catalunya (IEEC), E-08034 Barcelona, Spain \and
Department of Astronomy, Oskar Klein Center, Stockholm University, SE-106 91 Stockholm, Sweden \and
UCO/Lick Observatory, PO Box 85, Mount Hamilton, CA 95140, USA \and
Institute of Cosmology and Gravitation, University of Portsmouth, Dennis Sciama Building, Burnaby Road, Portsmouth PO1 3FX, UK \and
Department of Astrophysics, American Museum of Natural History, Central Park West and 79th Street, New York, NY 10024-5192, USA \and
Astronomical Observatory, University of Warsaw, Al. Ujazdowskie 4, 00-478 Warszawa, Poland \and 
The NSF AI Institute for Artificial Intelligence and Fundamental Interactions, USA \and
Department of Physics and Astronomy, University of California, Davis, 1 Shields Avenue, Davis, CA 95616-5270, USA \and
Department of Physics, Oskar Klein Centre, Stockholm University, SE-106 91, Stockholm, Sweden \and
Instituto de Astrofísica, Pontificia Universidad Católica de Chile, Av. Vicuña MacKenna 4860, 7820436, Santiago, Chile \and
Department of Astronomy, University of Virginia, 530 McCormick Rd, Charlottesville, VA 22904, USA \and
Tuorla Observatory, Department of Physics and Astronomy, FI-20014 University of Turku, Finland \and
Finnish Centre for Astronomy with ESO (FINCA), FI-20014 University of Turku, Finland \and
Department of Physics and Astronomy, Northwestern University, 2145 Sheridan Rd, Evanston, IL 60208, USA \and
Department of Astronomy, Kyoto University, Kitashirakawa-Oiwake-cho, Sakyo-ku, Kyoto, 606-8502. Japan \and
School of Physics, Trinity College Dublin, The University of Dublin, Dublin 2, Ireland \and
Research School of Astronomy and Astrophysics, Australian National University, Canberra, ACT 2611, Australia \and
University of California Observatories, 1156 High Street, Santa Cruz, CA 95064, USA \and
NSF-Simons AI Institute for the Sky (SkAI), 172 E. Chestnut St., Chicago, IL 60611, USA \and
Aryabhatta Research Institute of Observational Sciences (ARIES), Manora Peak, Nainital - 263001, India \and
Instituto de Ciencias Exactas y Naturales (ICEN), Universidad Arturo Prat, Chile \and
Department of Astronomy, The University of Texas at Austin, 2515 Speedway, Stop C1400, Austin, TX 78712, USA \and
Department of Physics and Astronomy, The Johns Hopkins University, 3400 North Charles Street, Baltimore, MD 21218, USA \and
Observatories of the Carnegie Institute for Science, 813 Santa Barbara Street, Pasadena, CA 91101-1232, USA \and
Department of Physics and Astronomy, Purdue University, 525 Northwestern Avenue, West Lafayette, IN 47907-2036, USA \and
School of Astronomy and Space Science, University of Chinese Academy of Sciences, Beijing 100049, People's Republic of China \and
National Astronomical Observatories, Chinese Academy of Sciences, Beijing 100101, People's Republic of China \and
Department of Astronomy and Astrophysics, University of Chicago, William Eckhart Research Center, 5640 South Ellis Avenue, Chicago, IL 60637, USA \and
Astrophysics sub-Department, Department of Physics, University of Oxford, Keble Road, Oxford, OX1 3RH, UK \and
MTA-ELTE Lend\"ulet "Momentum" Milky Way Research Group, Hungary \and
National Astronomical Research Institute of Thailand, 260 Moo 4, Donkaew, Maerim, Chiang Mai 50180, Thailand \and
Konkoly Observatory, HUN-REN Research Centre for Astronomy and Earth Sciences, MTA Centre of Excellence, Konkoly Thege Mikl\'os út 15-17, Budapest, 1121, Hungary \and
ELTE E\"otv\"os Lor\'and University, Institute of Physics and Astronomy, P\'azm\'any P\'eter s\'et\'any 1/A, Budapest, 1117 Hungary \and
Physics Department and Tsinghua Center for Astrophysics, Tsinghua University, Beijing, 100084, People's Republic of China
}


   \date{Received September 30, 20XX}


  \abstract
   {We present an abundance tomography analysis of SN 2024pxl, an intermediate luminosity Type Iax supernova ($M_\mathrm{r}=-16.82 \pm 0.19$ mag), with the most-detailed follow-up in the Type Iax subclass to-date. As one of the few intermediate luminosity Type Iax objects, SN 2024pxl may link the two extremes of the peculiar thermonuclear supernova subclass. To test this hypothesis, we analyze its spectral evolution through the first 100 days after the explosion and aim to probe the structure of its ejecta. We conduct an abundance tomography analysis using synthetic spectra produced with the one dimensional radiative transfer code TARDIS. The fit of the spectral time series provides a radial scan of physical properties and probes the stratification of chemical elements throughout most of the SN ejecta. The observed spectral evolution is well fit with the final model, similar to the general predictions of the pure deflagration scenarios, but significant modifications are required in the density function of the inner ejecta and in the chemical profiles of the outermost regions. The constrained physical characteristics, such as the photospheric velocities and the time of maximum light, are also consistent with other SNe~Iax. Despite their small number, intermediate luminosity Type Iax SNe are not outliers in the subclass but demonstrate the continuous nature of SNe Iax through their luminosity range. Following this observation, we argue that all SNe~Iax share the same progenitor and explosion origin.}

   \maketitle

\section{Introduction}

Type Iax supernovae (SNe Iax) form one of the most numerous subclasses of white dwarf (WD) explosions. Their estimated rate for all WD SNe varies between 5 and 30\% based on different volume-limited samples \citep[][respectively]{Dimitriadis24,Foley13}. The main properties of SNe Iax are their relatively low-luminosity, along with other peculiar characteristics of a low-energy thermonuclear explosion. Unlike other subclasses of SNe Ia, the Iax group exhibits great diversity in all of these properties. The peak absolute magnitudes range from $-18.6$ mag \citep[SN 2011ay;][]{Szalai15} to $-12.6$ mag \citep[SN 2021fcg;][]{Karambelkar21} in the r/R-band, while the expansion velocities of the photosphere span from $\sim$2000 \citep[SN 2008ha;][]{Foley09} to $\sim$9000 km\,s$^{-1}$ \citep[SN 2012Z;][]{Stritzinger15} at the moment of maximum light. 
Spectroscopically, SNe Iax are similar to ``normal'' SNe Ia at early epochs, although the well-known features of intermediate mass elements (IMEs), such as the Ca II H\&K or the Si II $\lambda6355$ lines, are less prominent. At the same time, features of iron group elements (IGEs), such as the regions around 4100 and 5200 \r{A} formed by Fe II, are dominant from the first days indicating the presence of IGEs in the outer layers of the SN ejecta. SNe Iax never turn fully nebular at later epochs; instead, they show permitted P Cygni and forbidden emission lines simultaneously, even years after the explosion \citep{Camacho-Neves23}. The blackbody emitting source required by this phenomenon can be explained via a bound remnant, which is formed from the burned material falling back due to the insufficient kinetic energy that failed to fully unbind the progenitor WD.

The most promising explosion scenario, the so-called pure or ``failed'' deflagration \citep{Jordan12,Kromer13} can explain both the relatively low luminosities and the low kinetic energies. All published hydrodynamic models \citep{Jordan12,Kromer13,Long14,Fink14,Kromer15,Leung20,Lach22} predict spherically symmetric ejecta, with a uniform abundance structure due to heavy mixing. The nearly constant mass fractions of chemical elements, dominated by unburned material (C and O from the WD) and IGEs, can successfully reproduce the observed spectral evolution in general \citep[see e.g.][]{Kwok25} around and after maximum (i.e., over the majority of the ejecta), even at extremely late epochs, which probe the innermost regions of SNe Iax \citep{Camacho-Neves23}. The assumption of completely uniform ejecta has been questioned by both the locations of line forming regions \citep{Stritzinger15} and abundance tomography analyzes \citep{Barna18,Barna21a,Singh23,Singh24,Barna26}. However, the signs of a stratified abundance structure are limited to early epochs, and thus, the outer regions of the ejecta. 

State-of-the-art pure deflagration models of C/O WDs can only account for the relatively luminous (RL, with a peak absolute magnitude of $M_\mathrm{r}<-17$ mag) and intermediate luminosity (IL, with $-17 < M_\mathrm{r}< -14.5$ mag) cases of the SN Iax class. For the extremely low-luminosity group \citep[EL, $M_\mathrm{r} > -14.5$][]{Barna26}, the only plausible solution so far has been the deflagration of a hybrid C/O/Ne WD \citep{Meng14}. \cite{Kromer15} presented hydrodynamic simulations that reproduced the main observables of the prototype object SN 2008ha \citep[$M_\mathrm{R}=-14.3$,][]{Foley09}. The lack of a general explanation for the explosion scenario of the EL sample, including the even less luminous objects discovered since SN 2008ha \citep[e.g. SNe 2019gsc, 2021fcg, 2024vjm;][]{Tomasella20,Srivastav20,Karambelkar21,Zimmerman26}, has raised the question of whether SNe Iax indeed form a few-parameter family \citep{Barna18} or can be distinguished into two (or more) subclasses with distinct observables and/or origins \citep{Singh23}. To resolve this question, the slowly growing IL sample is of great importance, as these transitional objects can constrain the continuous nature of SNe Iax over the entire luminosity range.

SN 2024pxl is only the third IL SN Iax with a multi-wavelength photometric dataset and spectral series, after SNe 2019muj \citep{Barna21a} and 2022xlp \citep{Banhidi25}. The study of these objects may link the two extremes of SNe Iax, assuming that the physical properties change continuously through the luminosity range of the subclass. The follow-up of SN 2024pxl provided the most detailed dataset of the subclass, both in the covered time period and wavelength range. The monitoring of the SN started just a few days after the explosion and continued for an unprecedented late +500 days. While it was observed by Swift, the Hubble Space Telescope, and numerous ground-based telescopes, SN 2024pxl was also the first SN Iax that was targeted by the James Webb Space telescope with NIRSpec and MIRI detectors \citep{Kwok25}. The analysis of the emission lines at near- and mid-infrared wavelengths (NIR and MIR) provided constraints on the symmetry and uniformity of the ejecta. All these factors made SN 2024pxl the most well-observed and, simultaneously, one of the most important objects in the Iax subclass. After detailed photometric and spectroscopic studies \citep{Hoogendam25,Kwok25,Singh26}, we aim to reconstruct the properties of the ejecta via abundance tomography to complete its analysis.

The paper is organized as follows. Section \ref{sec:sn2024pxl} summarizes the published results regarding SN 2024pxl and the observed datasets. Section \ref{sec:tardis} presents the abundance tomography, including the description of the radiative transfer code TARDIS and the adopted fitting strategy. The results of the analysis are presented in Section \ref{sec:results}, while the main conclusions are summarized in Section \ref{sec:conclusions}.\\

\section{SN 2024pxl}
\label{sec:sn2024pxl}

SN 2024pxl was discovered by the \texttt{BTSbot} machine-learning model \citep{Rehemtulla24} at 60514.4 MJD using observations from the Zwicky Transient Facility \citep[ZTF,][]{Graham19}. Its host galaxy, NGC 6384, is a spiral galaxy at a redshift of $z=0.0056$, and has been the subject of several distance measurements over the last two decades. Most of these estimates were obtained via the Tully-Fisher method, with results varying between 18.7 and 29.7 Mpc. As the most recent estimate, \citet{Singh26} presented a light curve analysis of the Type Ia SN 2017dh, which exploded close to the center of NGC 6384. The well-sampled $BgVri$ photometry was fitted with the BayeSN code \citep{Thorp21,Mandel22} and constrained its distance to $d=23\pm2.0$ Mpc (assuming $H_\mathrm{0}=73$ km\,s$^{-1}\,$Mpc$^{-1}$). This value is consistent with both previous results and matches well with that of Cosmicflow-3 ($d=25.1\pm4.0$ Mpc), the latest Tully-Fisher study \citep{Tully16}.

\begin{figure}[h!]
   \centering
   \includegraphics[width=\columnwidth]{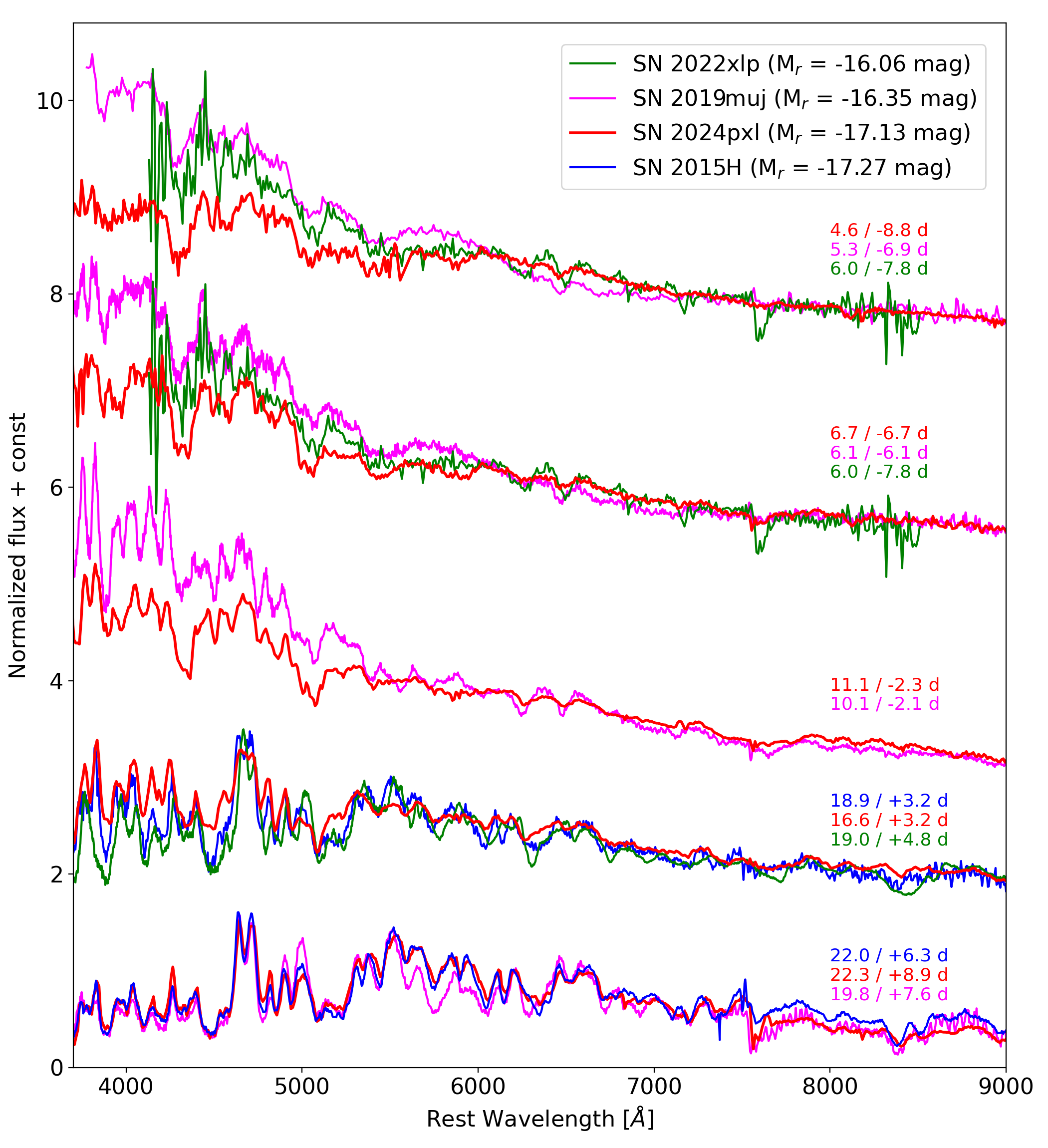}
      \caption{The comparison of early spectral evolution of SN 2024pxl (red) with that of a less luminous SNe 2019muj and 2022xlp \citep[][magenta and green, respectively]{Barna21a,Banhidi25} and the more luminous SN 2015H \citep[][blue]{Magee16} SNe Iax. The listed epochs show the time since explosion / the time since the $r$ band maximum in days.}
         \label{fig:spectral_comparison}
   \end{figure}

The observed spectra of SN 2024pxl suffered significant reddening. \cite{Schlafly11} estimated $E(B-V)_\mathrm{MW} = 0.11$\,mag as the Milky Way extinction in the direction of NGC 6384. The host galaxy reddening is constrained based on the narrow Na I D features by \cite{Singh26}, because the lines at the redshift of NGC 6384 have similar widths to those at Milky Way wavelengths. Thus, the total reddening is assumed to be $E(B-V) = 0.22$\,mag. Note that this value of extinction might still be underestimated, as the slope of the early spectral continuum falls short of that of SNe 2019muj and 2022xlp, two Type Iax SNe with similar peak absolute magnitudes (see Fig. \ref{fig:spectral_comparison}). However, the redder continuum of SN 2024pxl could be due to intrinsic properties as well.

    \begin{table}[!h]
      \caption{Observables and estimated light curve properties of SN 2024pxl adopted from \cite{Singh26}. } 
      \centering
        \label{tab:Singh26}
        \begin{tabular}{l | l } 
            \hline
            \noalign{\smallskip}
            Host Galaxy             &       NGC 6384 \\
            \noalign{\smallskip}
            \hline
            \noalign{\smallskip}
            redshift                &       0.0056 \\
            distance [Mpc]          &       23.0 $\pm$ 2.0 \\
            distance modulus [mag]  &       31.81 $\pm$ 0.11 \\
            E(B-V)$_\mathrm{MW}$ [mag]   &       0.11 \\
            E(B-V)$_\mathrm{host}$  [mag] &        0.11 \\
            \noalign{\smallskip}
            \hline
            \noalign{\smallskip}
            $T_\mathrm{first}$ [MJD]            &    $60513.9 \pm 0.1$ \\
            $T_\mathrm{max}$ in V-band [MJD]    &    $60526.9 \pm 0.5$ \\
            $T_\mathrm{max}$ in r-band [MJD]    &    $60527.6 \pm 0.5$ \\
            $M_\mathrm{V}(max)$ [mag]           &    $-16.58 \pm 0.19$ \\
            $M_\mathrm{r}(max)$ [mag]           &    $-16.82 \pm 0.19$ \\
            \noalign{\smallskip}
            \hline
         \end{tabular} 
\end{table}

The peak absolute magnitudes are $M_\mathrm{V}=-16.6$\,mag and $M_\mathrm{r}=-16.8$\,mag \citep{Singh26}, which place SN 2024pxl in the intermediate luminous (IL) group of the Type Iax class. Due to its transitional luminosity and proximity, SN 2024pxl became an ideal target for a follow-up campaign. Right after its discovery, high-cadence optical monitoring commenced, resulting in the most detailed photometric and spectroscopic dataset of a Type Iax SN. The ground-based data were collected and published by \cite{Singh26}, where the authors performed light curve analysis (including the estimation of peak properties, the time of first light, and decline rates), constructed and studied the bolometric light curve, and compared the spectral evolution of SN 2024pxl to that of other SNe Iax. In Table \ref{tab:Singh26} we summarize the main observables of SN 2024pxl as published by \cite{Singh26}.

SN 2024pxl was selected for target of opportunity observations with the James Webb Space Telescope and became the first Type Iax SN observed with the Near-Infrared Spectrograph (NIRSpec) and the Mid-Infrared Instrument (MIRI). The four spectral epochs presented by \cite{Kwok25} sampled the panchromatic evolution between +11 and +40 days with respect to $r$ band maximum, allowing for the study of the simultaneously present photospheric and nebular lines. The analysis of the forbidden lines showed similar and centrally symmetric line profiles for all identified elements, indicating a uniform chemical distribution.

In this study, we use the published spectral series of \cite{Singh26} and also adopt the estimated date of maximum light ($T_\mathrm{max} = 60527.6 \pm 0.5$ MJD in the $r$ band) as the reference time. The detailed description and log of observations can be found in \cite{Singh26}; here, Table \ref{tab:spectra} only lists those spectra that were adopted for spectral synthesis in our study. One spectrum is obtained with the Double Beam Spectrograph \citep[DBSP;][]{Oke82} from Palomar Observatory, with the Robert Stobie Spectrograph (RSS) of the Southern African Large Telescope (SALT), with the Kast double spectrograph at Lick Observatory, and with the Low Resolution Imaging Spectrometer \citep[LRIS;][]{Oke95} on the Keck I telescope. Two epochs were observed with the Alhambra Faint Object Spectrograph and Camera (ALFOSC) mounted on the Nordic Optical Telescope (NOT), and four epochs were acquired through the Global Supernova Project (GSP) collaboration using the telescopes of Las Cumbres Observatory \citep[LCO;][]{Brown13}. These spectra are chosen to sample the ejecta almost uniformly: the time intervals are shorter at earlier epochs when the photospheric velocity is expected to decrease rapidly, while the sampling becomes less frequent when the spectral evolution slows down.  Additionally, two yet unpublished spectra are also involved in the analysis, which were obtained by the ePESSTO+ project \citep{Smartt15} with the ESO Faint Object Spectrograph and Camera version 2 \citep[EFOSC2;][]{Buzzoni84} at the 3.6-m New Technology Telescope (NTT) are used here. 
The epochs of SN 2024pxl chosen for this study sample the first 100 days after the explosion. All spectra are scaled to photometry using polynomial functions and dereddened for the total assumed reddening of $E(B-V) = 0.22$ mag.

    \begin{table*}[!h]
      \caption{Log of the spectra of SN 2024pxl involved in the abundance tomography analysis. } 
      \centering
        \label{tab:spectra}
        \begin{tabular}{c c c c c c} 
            \hline
            \noalign{\smallskip}
            \textbf{Date} & \textbf{MJD} & \textbf{t$_\mathrm{exp}$ [d]} & \textbf{Phase [d]} & \textbf{Telescope/Instrument} & \textbf{Wavelength range [\r{A}]} \\
            \noalign{\smallskip}
            \hline
            \noalign{\smallskip}
            2024-07-24 & 60515.34 & 4.6 & $-$8.8 & P200/DBSP & 3400 - 10500 \\
            2024-07-26 & 60517.45 & 6.7 & $-$6.7 & FTN/FLOYDS & 3500 - 10000 \\
            2024-07-28 & 60519.26 & 8.5 & $-$4.9 & FTN/FLOYDS & 3500 - 10000 \\
            2024-07-30 & 60521.79 & 11.1 & $-$2.3 & SALT/RSS & 3500 - 9400 \\
            2024-08-02 & 60524.69 & 14.0 & $+$0.6 & Lick/Kast & 3600 - 10700 \\
            2024-08-05 & 60527.29 & 16.6 & $+$3.2 & FTN/FLOYDS & 3500 - 10000 \\
            2024-08-10 & 60532.97 & 22.3 & $+$8.9 & NOT/ALFOSC & 3500 - 9800 \\
            2024-08-18 & 60540.32 & 29.6 & $+$16.2 & FTN/FLOYD & 3500 - 10000 \\
            2024-08-31 & 60553.04 & 42.3 & $+$28.9 & NTT/EFOSC2 & 3650 - 9250 \\
            2024-09-08 & 60561.05 & 50.3 & $+$36.9 & NTT/EFOSC2 & 3650 - 9250 \\
            2024-09-21 & 60574.86 & 64.1 & $+$50.75 & NOT/ALFOSC & 3500 - 9650 \\
            2024-10-30 & 60613.20 & 103.5 & $+$89.09 & Keck/LRIS & 3650 - 10000 \\
            \noalign{\smallskip}
         \end{tabular} 
             \tablefoot{$t_\mathrm{exp}$ shows the time since the date of explosion constrained in the abundance tomography (MJD 60510.9); while the phases are given relative to the maximum in r-band (MJD 60527.6).}
\end{table*}

\section{Abundance tomography}
\label{sec:tardis}

Abundance tomography analysis is a powerful technique for studying expanding ejecta. As the SN expands, the photosphere recedes, allowing the observer to see deeper into the ejecta. The new layers contributing to the spectrum are denser and thus predominantly affect the formation of the spectral lines at a certain epoch. By fitting the spectral time series with the same ejecta model, we can map both the physical properties and the abundance of line forming elements layer by layer.

Since the first abundance tomography study of SN 2002bo \citep{Stehle05}, only a handful of objects have been analyzed using this method. Mapping the distribution of the chemical elements provided an effective tool to distinguish between the various explosion scenarios; thus, tomography analysis was adopted for multiple thermonuclear SNe: the well-observed normal Type Ia SNe 2011fe \citep{Mazzali14,Mazzali15} and 2014J \citep{Ashall14}, the transitional Type Ia SN 2021rhu \citep{Harvey23}, the subluminous SN 1986G \citep{Ashall16}, the super-Chandra SN 2009dc \citep{Hachinger12}, and 1991T-like SNe \citep{Sasdelli14,Obrien24}. The method was also used to constrain the IGE abundances in outer layers, which have a particular impact on the observed UV fluxes \citep{Hachinger13,Barna21b}. However, the broad and often saturated absorption features of SNe Ia increase the degeneracy of spectral fits and thus limit the accuracy of the inferred ejecta models. Type Iax supernovae exhibit more numerous and narrower spectral lines, significantly increasing both the number and quality of the fitting constraints. To date, 16 of these subluminous events, $\sim$13\% of the discovered Iax population, have been the subject of abundance tomography analysis. The full list and references of the objects can be found in Table \ref{tab:references}. All of the spectral synthesis of these tomography analyzes was performed with the 
1D Monte Carlo-based radiative transfer code TARDIS \citep{Kerzendorf14}.

TARDIS provides robust tracking of photon propagation in a homologously expanding atmosphere, which allows for obtaining reliable ejecta properties even with the computational power of a desktop PC. The code also allows for the design of SN ejecta by defining the densities and chemical composition of radial layers; thus, it provides a powerful tool to synthesize and fit SN spectra. Here we use the v2025.10.12 release of the radiative transfer code TARDIS \citep{tardis25}. The code assumes radially symmetric ejecta with a sharp blackbody emitting photosphere and follows the propagation of photon packets through the predefined layers via Monte Carlo simulations. Multiple options are available for light-matter interaction, excitation, and ionization rates. The chosen options and the convergence strategy of the TARDIS model runs are presented in Appendix \ref{tab:TARDIS-settings}, while the fitting strategy is described in Sec. \ref{sec:tardis}.

In this study, we aim to constrain the mass fraction of twelve chemical elements, assuming a non-uniform distribution. We chose the seven most abundant elements (O, C, Si, S, Fe, Co, Ni) of the \textit{N1def} model and five additional elements (namely Ca, Mg, Ti, V, Cr), whose line formation impact has been previously identified in Type Iax spectra by abundance tomography studies. The abundance of every other chemical element is fixed at the same value as in the N1def model \citep{Fink14}, which has been shown to provide a solid match with the spectral features of SN 2024pxl \citep{Kwok25}. In the case of mass fraction deficit, we choose oxygen as the ``filler'' element for normalization, since its spectral lines are well distinguished and the overestimation of its mass fraction does not spoil the fits. We define radial layers with velocity steps of 1000 km\,s$^{-1}$, in which we fit the mass fraction of the chemical elements. This choice is necessary to test the possible stratification of the ejecta, but it provides a large number of free parameters, which is unsuitable for exploring the entire parameter space. Instead, we set all mass fractions to a constant value representing the general abundances of the N1def model as an initial guess and alter the mass fraction of any layers if the fit of a spectral feature requires it. Note that this fitting strategy is not aimed at finding the best-fit synthetic spectra (even though we use the term ``best-matching'' in the following for simplification); rather, it is intended to provide a plausible solution for the ejecta model.

\begin{figure}

	\includegraphics[width=\columnwidth]{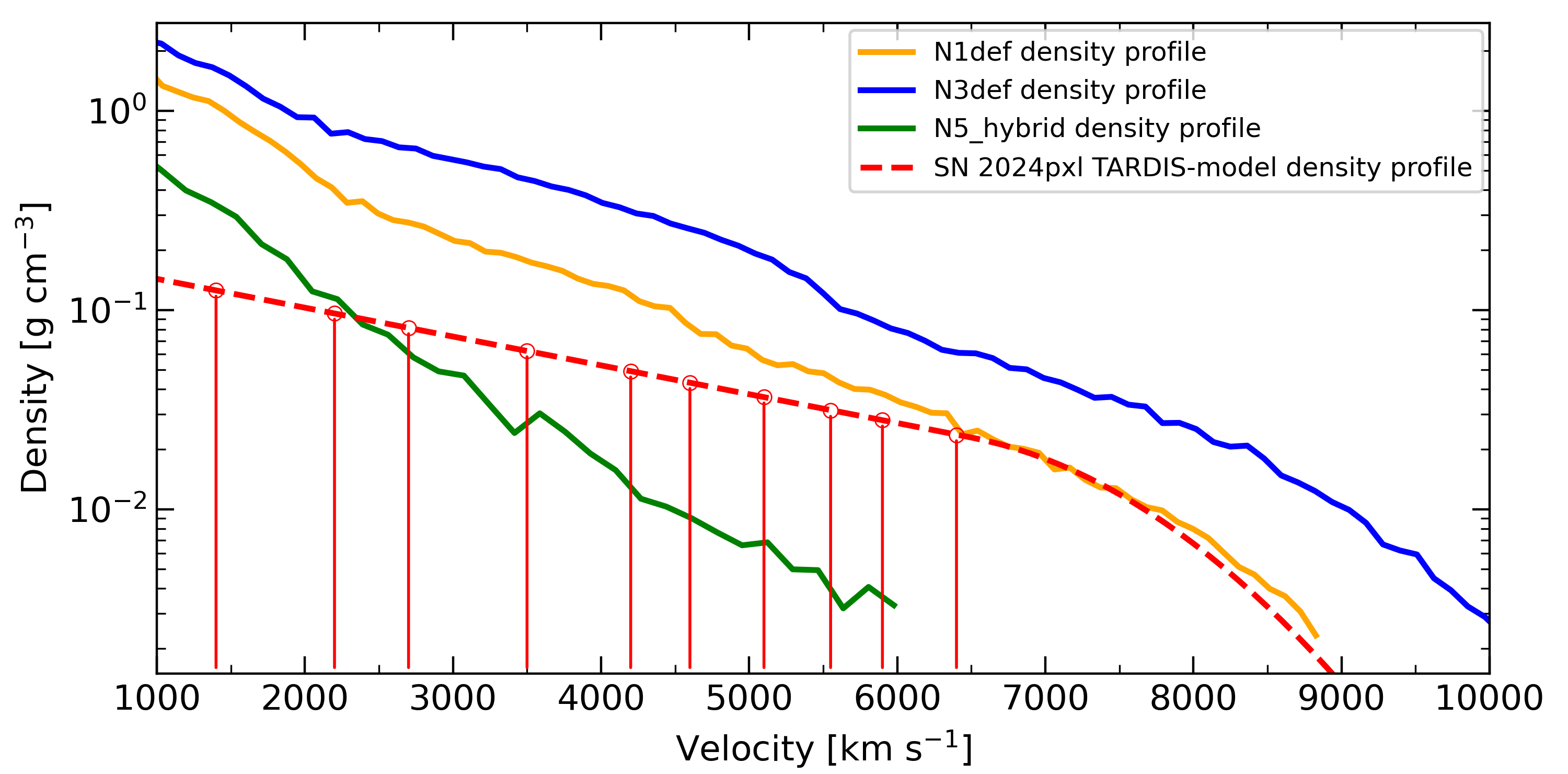}
    \caption{The assumed density function of the best-matching TARDIS model for SN 2024pxl (red). The density structures of the pure deflagration models with similar peak abosulte magnitudes as SN 2024pxl are also plotted for comparison. The vertical lines show the velocity of the photosphere at the studied epochs. }
    \label{fig:densities}
\end{figure}

We adopt a density formula similar to that of \cite{Barna21a}, which contains an exponential inner part and a shallow cut-off toward higher velocities:
\begin{equation}
\rho (v,t_{\mathrm{exp}}) = 
\begin{cases}
\rho_0 \cdot \left(\frac{t_{\mathrm{exp}}}{t_0}\right)^{-3} \cdot \exp\left({-\frac{v}{v_0}}\right)
& \text{for } v \leq v_\mathrm{cut} \\
\rho_0 \cdot \left(\frac{t_{\mathrm{exp}}}{t_0}\right)^{-3} \cdot \exp\left({-\frac{v}{v_0}}\right) \cdot
C^{-\frac{(v - v_\mathrm{cut})^2}{v^{2}_\mathrm{cut}}}
& \text{for } v > v_\mathrm{cut}
\end{cases}
\label{eq:density}
\end{equation}  

where $v$ is the velocity coordinate, $\rho_\mathrm{0}$ is the core density (i.e., density at $v=0$ km\,s$^{-1}$) at the reference time $t_\mathrm{0}$; $v_\mathrm{0}$ defines the decline rate of the exponential density profile; $v_\mathrm{cut}$ marks the start of the rapid decline of the outer regions, and $C$ scales the strength of the cut-off. Eq. \ref{eq:density} can effectively reproduce the density profiles of pure deflagration models (for an example, see Fig. \ref{fig:densities}). To reduce the number of free parameters, we fix the strength of the cut-off as $C = 8.2 \times 10^{5}$, which produces a similar decline in the density function to that of the N1def model. The $v_\mathrm{cut}$ can be adjusted by the widths of the strongest absorption features, such as the Fe II lines between 4500 and 5200 \r{A} both before and after maximum light. Fitting the blue wings of these lines indicates a velocity shift of $\sim$6500 km\,s$^{-1}$, which is adopted for $v_\mathrm{cut}$ in the whole abundance tomography. Note that during the abundance tomography, the core of the ejecta is not covered; thus, $\rho_\mathrm{0}$ cannot be constrained directly, and it acts as a simple scaling factor for the density profile. Moreover, the $\rho_\mathrm{0}$ and $v_\mathrm{0}$ parameters are strongly coupled because a spectral epoch samples only a short range of the whole ejecta, and the layers have continuously lower impact outward in the SN atmosphere. To handle the degeneracy between the parameters of Eq. \ref{eq:density}, we fix the slope of the exponential function to $v_\mathrm{0}=3000$ km\,s$^{-1}$, which has already provided plausible solutions for the spectral synthesis of similar Type Iax SNe such as SNe 2015H and 2022xlp \citep[$M_\mathrm{r}=-17.27$ and $-16.06$,][respectively]{Magee16,Banhidi25}.
The date of the explosion is fit by assuming that the moment of first-light \citep[$T_\mathrm{first} = 60513.86 \pm 0.10$ MJD,][]{Singh26} adopted from the LC analysis works as an upper limit. Luminosities and photospheric velocities are constrained individually for each spectrum.

\section{Results}
\label{sec:results}

\begin{figure}
   \centering
   \includegraphics[width=\columnwidth]{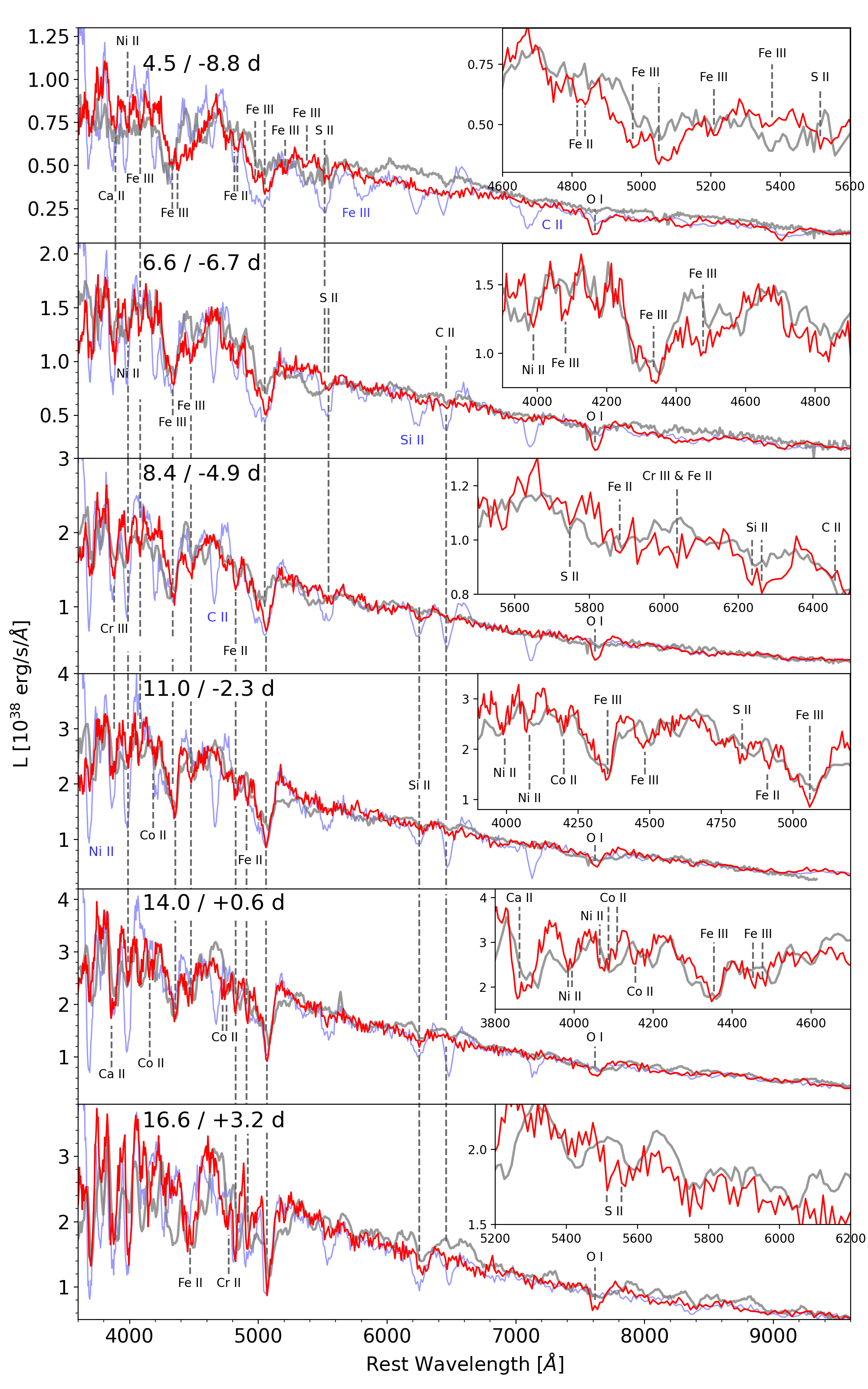}
      \caption{Early spectral evolution of SN 2024pxl (grey) fitted with the TARDIS synthetic spectra (red) of the the abundance tomography analysis. The unambiguous identification of spectral features based on the TARDIS photon packet statistics are marked with dashed lines. TARDIS spectra calculated by assuming the N1def abundance structure (blue) are also shown for comparison.}
         \label{fig:tardis1}
   \end{figure}

\begin{figure}
   \centering
   \includegraphics[width=\columnwidth]{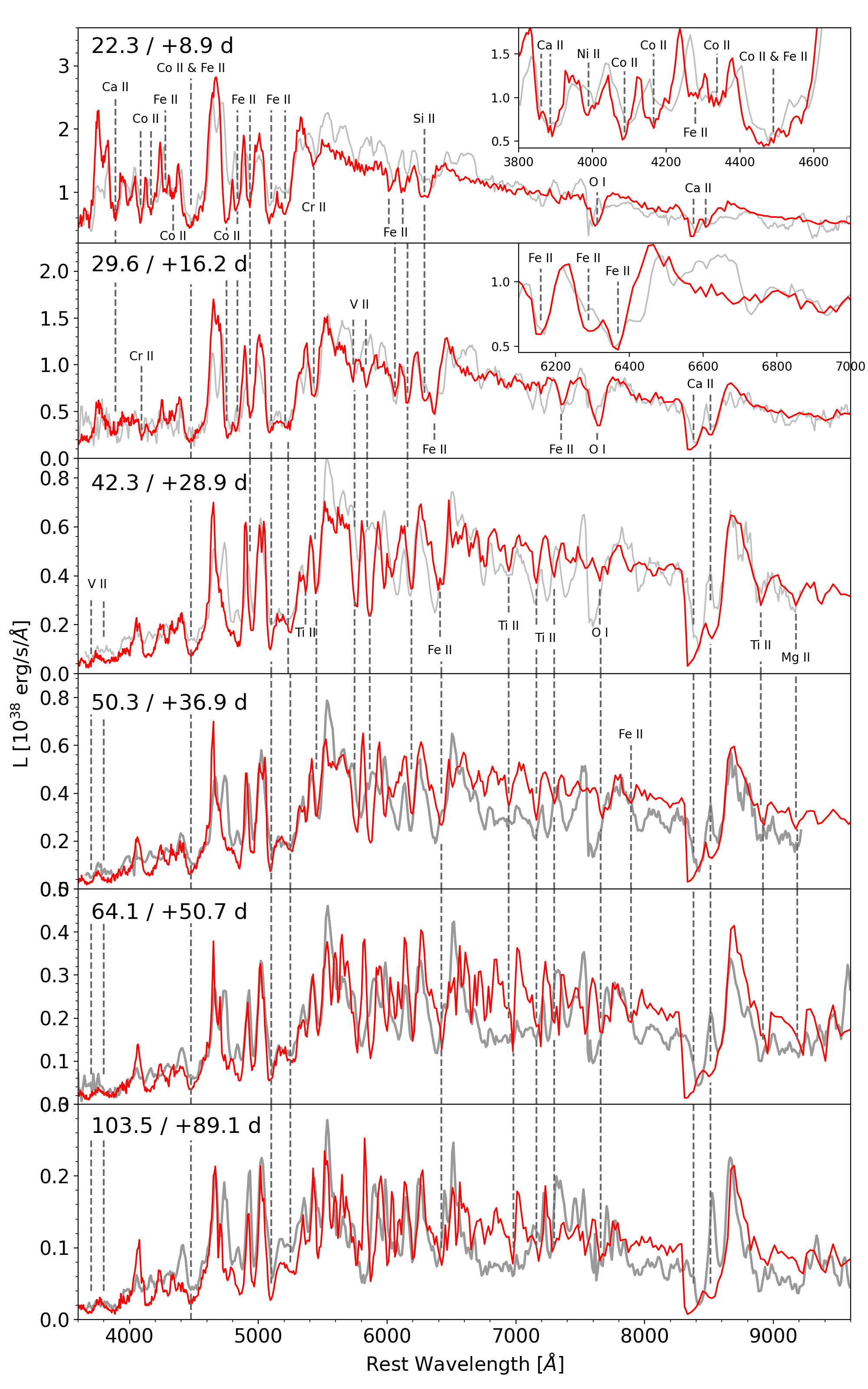}
      \caption{Late spectral evolution of SN 2024pxl (grey) fitted with the TARDIS synthetic spectra (red) of the the abundance tomography analysis.}
         \label{fig:tardis2}
   \end{figure}

The TARDIS radiative transfer code and the fitting strategy described in Sec. \ref{sec:tardis} were proven to be effective methods for synthesizing the spectral evolution of SN 2024pxl. TARDIS becomes less reliable with time after maximum light for normal SNe Ia because it works in the optically thick regime. However, it can be used for SNe Iax even at late epochs due to the existence of a continuum emitting photosphere possibly originating from a bound remnant \citep{Foley16, Camacho-Neves23}. As presented in Figs. \ref{fig:tardis1} and \ref{fig:tardis2}, both the continuum in general and most of the P Cygni lines are reproduced with the same ejecta model over 100 days of evolution.  

The evolved spectral features of the first spectrum (60515.34 MJD; $t_\mathrm{exp}=4.6$ days), such as the two Fe II dominated regions around 4300 and 5400 \r{A}, and the lack of a visible Si II $\lambda6345$ line, indicate that the photosphere has already reached the denser region of the ejecta but is still relatively hot, as assumed by the required excitation temperatures. These characteristics constrain the epoch to 4-6 days after the explosion, according to the analysis of other SNe Iax. The direct comparison of SN 2024pxl to its closest relatives by luminosity, SNe 2015H and 2019muj (see Fig. \ref{fig:spectral_comparison}), by peak luminosity ($M_\mathrm{r}=-17.27$ and $-16.35$ mag, respectively), also supports this estimation of the time since explosion ($t_\mathrm{exp}$). 

The density structure of the N1def model used as an initial guess had to be reduced, as the fits of lines at near- and post-maximum epochs required lower densities by an order of magnitude. The final model was constructed with a core density of $\rho_\mathrm{0}=0.2$ g\,cm$^{-3}$ (at the reference time of $t_\mathrm{exp}=100$ s), determined by the form of Eq. \ref{eq:density}. This produces densities close to the photosphere (in the most impactful line forming region) at the latest epochs that were an order of magnitude lower than those of the N1def model. These modifications also allowed for a realistic result for the date of explosion, which, based on the fits of the first two epochs, is constrained to $T_\mathrm{exp}=60510.7$ MJD with an approximate uncertainty of 1.5 days. Although a "dark phase” \citep{Piro13,Piro14} between the moment of explosion and the first observed light ($T_\mathrm{first}=60513.9$ MJD) is expected, the difference of $\sim3.2$ days marks a significant discrepancy.

The assumption of a sharp photosphere is robust throughout the studied time period. The location of the inner boundary of the ejecta, i.e., the velocity of the photosphere ($v_\mathrm{phot}$), affects not only the blueshift of multiple P Cygni features but also has a significant impact on the temperature profile of the SN atmosphere and, thus, on the ionization/excitation relations. More specifically, the fit of the thinner lines (such as Ni II $\lambda4068$, the S II W-feature, C II $\lambda6580$) and the relative strengths of Fe II and Fe III features between 4500-5200 \r{A} are effective ways to constrain $v_\mathrm{phot}$  before and around maximum light. At later epochs, $v_\mathrm{phot}$ was mainly estimated from modeling the shape of the continuum flux (via $T_\mathrm{phot}$) and the shift of several strong Fe II lines (see Fig. \ref{fig:tardis2}).

\begin{figure}

	\includegraphics[width=\columnwidth]{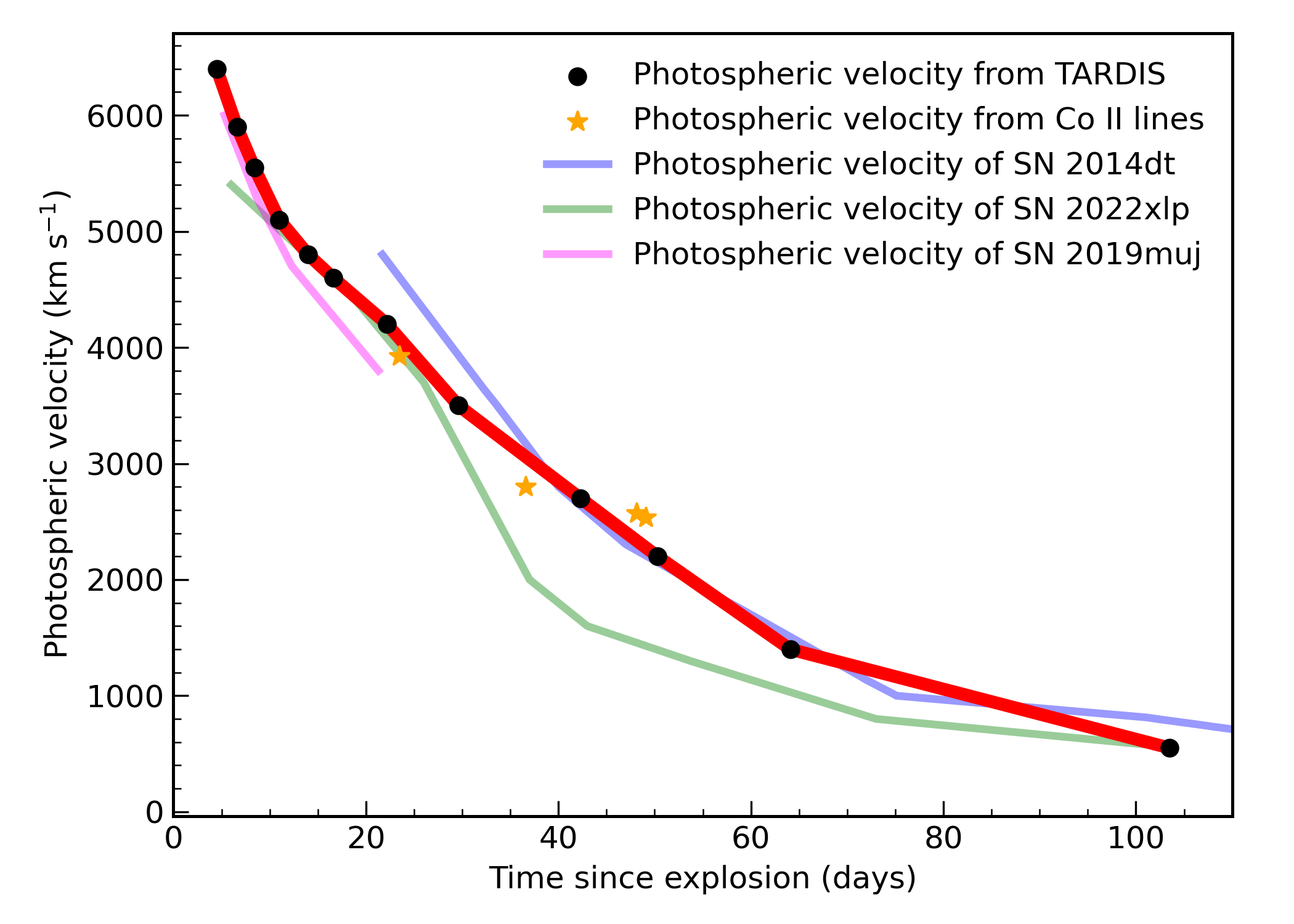}
    \caption{The photospheric velocity evolution of SN 2014pxl and other SNe Iax with similar absolute magnitude. }
    \label{fig:velocities}
\end{figure}

The complex impact on the synthetic spectrum allows $v_\mathrm{phot}$ to be set with sufficient precision of a few hundred km\,s$^{-1}$ in the abundance tomography analysis. The time function of $v_\mathrm{phot}$ (the inferred values are presented in Fig. \ref{fig:velocities}) declines rapidly at the earliest epochs, showing a nearly linear trend around maximum light, and slowly flattens during late-time. This resembles that of normal SNe Ia, with the difference being a non-ceasing photospheric phase.

\begin{figure}

	\includegraphics[width=\columnwidth]{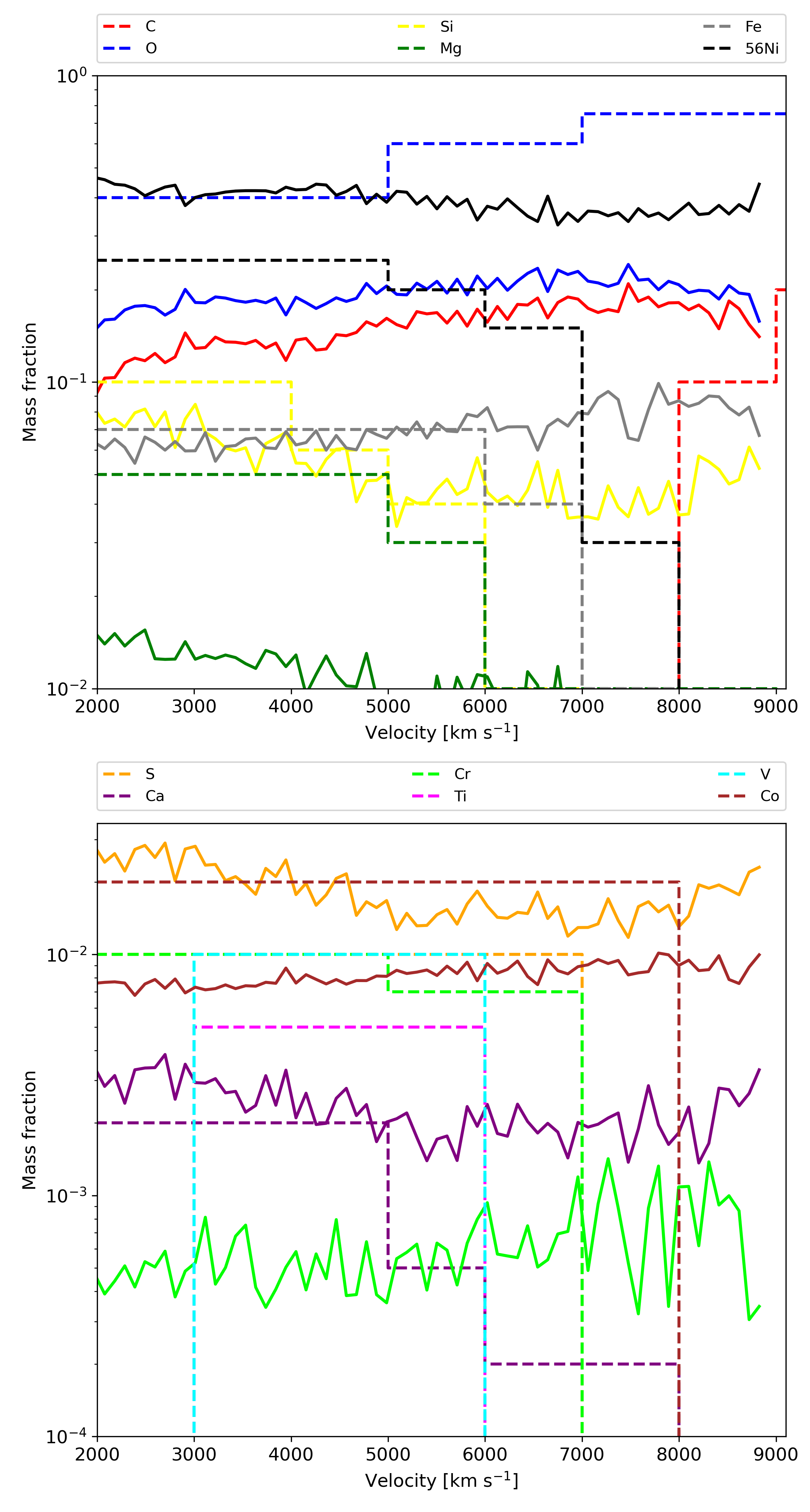}
    \caption{The mass fractions of chemical elements fitted in the abundance tomography analysis of SN 2024pxl (dashed lines), compared them to those of the N1def pure deflagration model (solid lines). }
    \label{fig:abundances}
\end{figure}

The best-matching chemical abundances are shown in Fig. \ref{fig:abundances}. Similar to other studies of SNe Iax, the outermost layers - sampled by the first epochs - cannot be described well with a uniform structure. As presented in Fig. \ref{fig:tardis1}, both C, IMEs (e.g., Si) and IGEs (e.g., Fe and $^{56}$Ni) would produce overwhelmingly strong and blueshifted lines if the initial abundance structure of the N1def model is kept. Instead, these elements are introduced in the best-matching model with increasing mass fractions in the deeper layers revealed at subsequent epochs. The outermost region over 6000 km\,s$^{-1}$ is dominated by C and O, representing the unburned material of the progenitor WD. While oxygen is used as a filler element in general, and thus its abundance is overestimated (see the misfit of the O I $\lambda 7779$ line), the carbon mass fraction is estimated according to the fit of C II $\lambda6580$ and is also constrained by the barely detectable C III $\lambda4647$ line. As a further discrepancy from the uniform abundance scheme of the deflagration model, carbon is limited to being present above 7000 km\,s$^{-1}$ in the final TARDIS model because even a few percentages of carbon mass fraction would form extremely strong features.

Around two weeks after maximum, the Fe II lines start to dominate the optical spectra. The only strong feature formed by another element is the Ca II NIR triplet, while weaker features can be clearly associated with magnesium, chromium, titanium, nickel, and oxygen. Other elements, such as silicon, sulfur, and carbon, do not show any identifiable lines, and the output spectra are relatively insensitive to their mass fractions. According to the fitting strategy described above, these elements are present with constant, N1def-like abundances in the deepest regions of our model.

The relationship between various physical properties and the peak luminosity of SNe Iax has been investigated, including the rise time, $\Delta m_\mathrm{15}$, and photospheric velocities. These studies could not identify any strong correlation, either because of the small sample or the uncertain properties, while the distribution of $\Delta m_\mathrm{15}$ parameters even raised the possibility of two populations within the Type Iax class \citep{Singh23}. In the case of the photospheric velocity, it can be difficult to accurately measure, due to numerous and  overlapping lines as compared to normal SNe~Ia. The classical method for thermonuclear SNe is based on the blueshift of the absorption minimum of the prominent and unblended Si II $\lambda6355$. For Type Iax SNe, the same feature appears significantly weaker, and it is polluted by Fe II $\lambda6456$ even a few days before maximum light. In a few days, the Fe II line dominates the observed P Cygni feature and makes the $v_\mathrm{phot}$ underestimated due to its longer rest wavelength \citep{Szalai15,Maguire23}. This effect provides an explanation for the scattering of the previously published luminosity-velocity relations and why the measured Doppler-shift of Si II is always lower than that of other elements.

In \cite{Banhidi25}, the authors suggested that abundance tomography analysis with radiative transfer codes could offer a better measurement for $v_\mathrm{phot}$. Reliable spectral synthesis was presented using radiative transfer codes with a sharp blackbody emitting inner ejecta boundary for thermonuclear SNe, implying that the assumption of the photosphere is robust. The naive $v_\mathrm{phot}$ estimate is based on the line forming region of chemical elements, which starts at the photosphere; thus, the Doppler-shift of the absorption minimum marks its location according to the formation of the P Cygni profiles. In radiative transfer codes like TARDIS, the location of the photosphere influences the density and temperature functions of the model ejecta, which greatly affects the shape of the continuum and the optical depth profiles of every line. Due to the numerous reconstructable lines, the location of the photosphere can be constrained with the precision of a few hundred km\,s$^{-1}$ at the early and near-maximum epochs. The uncertainty is mainly caused by the poorly constrained moment of explosion since $v_\mathrm{phot}$ is coupled with $t_\mathrm{exp}$. The uncertainty increases later as the strong, saturated features are less sensitive to the photosphere. Even at the latest epochs, a change of $v_\mathrm{phot}$ by 500 km\,s$^{-1}$ produced discrepancies that could not be counterbalanced by tuning other parameters. Thus, we can safely assume a general uncertainty of $\pm500$ km\,s$^{-1}$ for measurements at each epoch of SN 2024pxl.

The $v_\mathrm{phot}$ values inferred from the tomography analysis are shown in Fig. \ref{fig:velocities} as a function of $t_\mathrm{exp}$. The velocities estimated from the blueshifts of the observed series of the unblended NIR Co II lines \citep{Singh26} are also presented for comparison. The difference of $\sim200$ km\,s$^{-1}$ from the values of TARDIS fits between $t_\mathrm{exp}=20$ and 50 days indicates that the region forming the Co II line is indeed linked to the photosphere, providing an independent way to constrain $v_\mathrm{phot}$ and further validate the tomography method.

\subsection{Relation to other SNe Iax}

It has been shown in numerous studies that the spectroscopic evolution of SNe Iax is diverse \citep{Magee17, Dutta22}. However, such discrepancies might occur due to the physical properties of the ejecta instead of their chemical composition. In the case of more luminous explosions, higher ejecta masses have been estimated, which is consistent with the predictions of the deflagration models \citep{Fink14}. The differences are also present in light curves, as less energetic explosions tend to show more rapid evolution, although the correlation is less strong \citep[considering $\Delta m_\mathrm{15}$,][]{Singh23} or the estimates of light curve properties are uncertain \citep[rise time]{Magee25}.

In Fig. \ref{fig:spectral_comparison}, we compare the spectral evolution of SN 2024pxl with the slightly less energetic Type Iax SNe 2022xlp and 2019muj ($M_\mathrm{r}=-16.06$ and $-16.40$ mag, respectively), and the more energetic SN 2015H ($M_\mathrm{r}=-17.27$ mag), at similar epochs. The most obvious difference can be observed at pre-maximum epochs when SN 2019muj shows a steeper continuum flux, which can be interpreted as the result of a hotter photosphere, but could also be the consequence of the uncertain host-galaxy extinctions. At the same time, the observed set of lines is identical, and there are only moderate differences in the relative strength of the spectral features. The agreement between the spectra is even more striking after maximum, when the discrepancies in the continuum and P Cygni profiles almost completely disappear among the three objects. This behavior disfavors the assumption of additional host-galaxy extinction and suggests that SNe Iax with the same peak luminosity have very similar inner properties regarding their temperature, densities, and potentially the continuum source bound remnant. The differences are rather characteristic of the outer ejecta, which also actively shape the light curves before and around maximum light. 

In Fig. \ref{fig:velocities}, the photospheric velocities of SN 2024pxl are also compared to those of SNe Iax with either similar peak luminosity or a long coverage of spectral synthesis \citep{Camacho-Neves23, Banhidi25}. The photospheric evolutions show a similar trend in general: $v_\mathrm{phot}$ values rapidly decrease first, then the curves continuously flatten after $t_\mathrm{exp}$ $\simeq 50$ days. All three long-term velocity functions converge to a few hundred km\,s$^{-1}$. This feature is a potential clue for the existence of the super-Eddington wind emerging from the bound remnant predicted by the failed deflagration scenario \citep{Foley16}. Assuming a steady state wind, the $v_\mathrm{phot}=550$ km\,s$^{-1}$ and $\rho_\mathrm{phot}= 2.6 \times 10^{-1}$ g\,cm$^{-3}$ at 100 days indicate a mass-loss rate of $\sim 6 \times 10^{-4}$ M$_\odot$\,yr$^{-1}$. This is almost an order of magnitude lower than the mass loss rate of SN 2014dt \citep[$\sim 4 \times 10^{-3}$ M$_\odot$\,yr$^{-1}$,][]{Camacho-Neves23} estimated at a similar epoch.

\begin{figure}

	\includegraphics[width=\columnwidth]{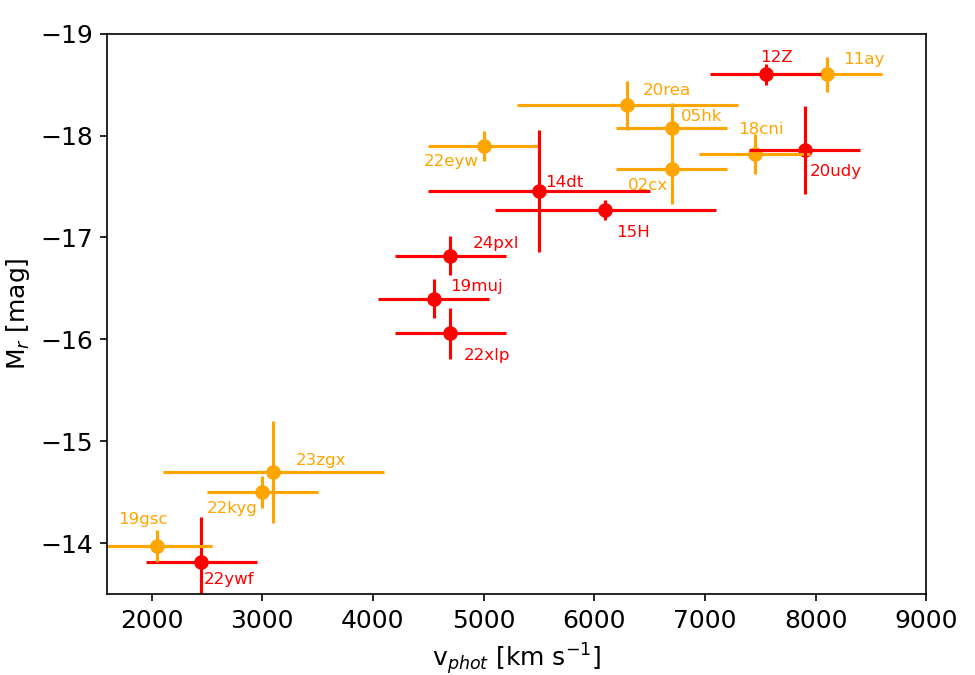}
    \caption{Absolute magnitudes in r/R-band of SNe Iax as a function of their photospheric velocity at the moment of maximum light.  We only plot SNe Iax which have been the subject of abundance tomography. The colors indicate wether the distance of the host galaxy was estimated from the Hubble-flow (orange) or based on any more precise indicators, such as Tully-Fisher relation or SNe (red). In case of no spectral coverage around the maximum (SNe 2014dt, 2015H, and 2023zgx), photospheric velocity is estimated with linear extrapolation and the approximate uncertainty is doubled. The corresponding references are listed in Tab. \ref{tab:references}.}
    \label{fig:vM}
\end{figure}

The interpolated photospheric velocity at the moment of the $r$ band maximum is 4700 km\,s$^{-1}$. This value fits perfectly into the trend of SNe Iax shown in Fig. \ref{fig:vM}, where the peak $r$ band absolute magnitudes are plotted as a function of $v_\mathrm{phot}$ values estimated for the moment of $r$ band maximum. Here we considered only those SNe Iax that have been the subject of abundance tomography analysis, as the naive method based on the blueshift of P Cygni features, typically that of Si II $\lambda6355$, usually underestimates $v_\mathrm{phot}$ \citep{Szalai15,Maguire23}. The continuity of the luminosity-velocity (L-v) correlation, supported by SN 2024pxl, is an indirect sign of the common origin of the whole Type Iax subclass. Note that the uncertainties of the adopted distances strongly affect the peak absolute magnitudes while also weakly affecting the spectral fits and thus the estimated $v_\mathrm{phot}$ values. For this reason, we marked those SNe in Fig. \ref{fig:vM} that were analyzed based on their relatively uncertain Hubble-flow distance. Due to this and the lack of total coverage over the luminosity range of SNe Iax, the proposed L-v correlation requires further validation.

\section{Conclusions}
\label{sec:conclusions}

We conducted an abundance tomography analysis on the type Iax SN 2024pxl, one of the most well-observed objects of its subclass. The peak luminosity increases the importance of SN 2024pxl, as it belongs to the small sample of the IL group, and as such, it may provide further insights into the link between the extremities of SNe Iax. The comparison with other SNe Iax with similar peak luminosity revealed practically the same set and shapes of spectral lines but significant differences in the continuum at pre-maximum epochs. This can be interpreted as intrinsic differences in the temperature and/or density profiles of the ejecta or as the result of the uncertain host-galaxy extinctions. Considering the similar inner ejecta properties inferred from the tomography analysis and the lack of spectral discrepancies around three weeks after the explosion, the source of different reddening is probably related to the outer regions of the ejecta.

The tomography analysis is carried out by spectral synthesis performed with the radiative transfer code TARDIS. The described fitting strategy ensures that, despite the large number of free parameters, we can effectively map the properties of the ejecta, including the possible layering of chemical elements. The analysis used the predictions of the single-ignition-point deflagration N1def model, which can effectively describe the ejecta structure of SN 2024pxl based on the line profiles of post-maximum NIR and MIR spectra \citet{Kwok25}, as an initial assumption for the model ejecta. Although the abundance and density structures of the N1def model produced the majority of the existing spectral features, the fit of their line profiles revealed several notable discrepancies. The updates to the abundance structure mainly affected the outermost layers, where both IMEs and IGEs are present with no or significantly decreased mass fractions in our best-fit model. This region has significantly lower densities than the rest of the ejecta; thus, it mainly affects the earliest spectra, while its impact on the forbidden emission line profiles is insignificant. The stratification of the inferred abundance structure indicates that the mixing of the ejecta after the explosion is not as strong as expected from current pure deflagration models.

The fits of the post-maximum optical spectra are consistent with the assumption that the ejecta are well mixed on large scales, as a constant mass fraction below 4000 km\,s$^{-1}$ provides the same quality of fitting as stratified solutions. Carbon was identified only in this outer region, but its presence between 4000 and 8000 km\,s$^{-1}$ would produce overwhelmingly strong carbon features in the early spectra. The carbon abundance is highly ambiguous below 4000 km\,s$^{-1}$, since no spectral features were identified during the post-maximum epochs. However, the N1def-like abundance of carbon in the inner region is also consistent with the spectral synthesis. As another significant difference from the predictions of N1def, the densities in the inner layers were decreased by an order of magnitude in our final ejecta model to fit the strong iron features dominating the optical wavelengths.

The photospheric velocity evolution of SN 2024pxl fits well into the sample of SNe Iax. The estimated velocity curves of the SNe show differences according to the absolute magnitudes at the early epochs, as the more energetic explosions expand faster, which provides a strong correlation between $v_\mathrm{phot}$ values and luminosities at the moment of maximum light. This attribute further supports the idea that SNe Iax originate from a similar explosion mechanism, as there are no breaks or outliers appearing in the IL-region of the velocity distribution. The $v_\mathrm{phot}$ values match well with those estimated from the Doppler-shift of Co II NIR features, instead of the classical method based on the Si II $\lambda6355$ feature \citep{Singh26}.


\clearpage
\begin{acknowledgements}
This research made use of \textsc{tardis}, a community-developed software package for spectral synthesis in supernovae \citep{Kerzendorf14}. The development of \textsc{tardis} received support from GitHub, the Google Summer of Code initiative, and from ESA's Summer of Code in Space program. \textsc{tardis} is a fiscally sponsored project of NumFOCUS. \textsc{tardis} makes extensive use of Astropy and Pyne. This work makes use of observations from the Las Cumbres Observatory network. The LCO team is supported by NSF grants AST-2308113 and AST-1911151. Based on observations collected at the European Organisation for Astronomical Research in the Southern Hemisphere, Chile, as part of ePESSTO+ (the advanced Public ESO Spectroscopic Survey for Transient Objects Survey – PI: Inserra). Based on observations collected at the European Southern Observatory under ESO programmes 112.25JQ.012 and 112.25JQ.010. B.B. received support from the Hungarian National Research, Development and Innovation Office grants OTKA PD-147091, the János Bolyai Research Scholarship of the Hungarian
Academy of Science and from the HUN-REN Hungarian Research Network. K.M., T.E.M.B. and J.H.T. acknowledge funding from Horizon Europe ERC grant no. 101125877.  Time-domain research by the University of Arizona team and D.J.S. is supported by National Science Foundation (NSF) grants 2108032, 2308181, 2407566, and 2432036. L.G. acknowledges financial support from CSIC, MCIN and AEI 10.13039/501100011033 under projects PID2023-151307NB-I00, PIE 20215AT016, CEX2020-001058-M, and by the MaX-CSIC Excellence Award MaX4-SOMMA-ICE. J.H.T. acknowledges Horizon Europe ERC grant no. 101125877. T.-W.C. acknowledges financial support from the Yushan Fellow Program of the Ministry of Education, Taiwan (MOE-111-YSFMS-0008-001-P1), and from the National Science and Technology Council, Taiwan (NSTC 114-2112-M-008-021-MY3). N.F. acknowledges support from the National Science Foundation Graduate Research Fellowship Program under Grant No. DGE-2137419. M.S. acknowledges financial support provided under the National Post Doctoral Fellowship (N-PDF; File Number: PDF/2023/002244) by the Science \& Engineering Research Board (SERB), Anusandhan National Research Foundation (ANRF), Government of India. KM acknowledges support from the Japan Society for the Promotion of Science (JSPS) KAKENHI grant (JP24KK0070, JP24H01810, and JP23H04894). Observations from coauthor A.J.M. were made under the aegis of the ASTRAL (Astronomy/STEM Alliance with Lick Observatory) consortium, supported by a generous grant from the Gordon and Betty Moore Foundation (PI B. Macintosh).

\end{acknowledgements}

\bibliography{example}
\bibliographystyle{aasjournal}

\onecolumn
\appendix
\section{Some extra material}

\begin{center}
\begin{longtable}{c | cc | cc} 
\caption{The peak absolute magnitudes in r/R-band and photospheric velocities of type Iax SNe at the moment of maximum light.} \label{tab:references} \\
\hline
\hline
\multicolumn{1}{c}{\textbf{Supernova}} & \multicolumn{2}{l}{\textbf{  $M_\mathrm{r}$ [mag]}} & \multicolumn{2}{l}{\textbf{  $v_\mathrm{phot}$ [km\,s$^{-1}$}]}  \\ \hline 
		SN 2011ay & -18.60 &  \citet{Szalai15} & 8100 & \citet{Barna17}\\
        SN 2012Z  & -18.60 &  \citet{Stritzinger15} & 7600 & \citet{Barna18}\\
        SN 2020rea & -18.30 & \citet{Singh22} & 6300 & \citet{Kayal25}\\
        SN 2005hk & -18.07 &  \citet{Stritzinger15} & 6700 & \citet{Barna18}\\
        SN 2022eyw & -17.90 & \citet{Hrishav26} & 5000 & \citet{Hrishav26}\\
        SN 2020udy & -17.86 &  \citet{Maguire23} & 7900 & \citet{Singh24}\\
        SN 2018cni & -17.82 &  \citet{Singh23} & 7500 & \citet{Singh23}\\
        SN 2002cx & -17.67 &  \citet{Li03} & 6700 & \citet{Barna18}\\
        SN 2014dt  & - 17.46 & \citet{Camacho-Neves23} & 5500 & \citet{Camacho-Neves23}\\
        SN 2015H & -17.27 &  \citet{Magee16} & 6100 & \citet{Barna18}\\
        SN 2024pxl & -17.13 & \citet{Singh26} & 4700 & This work\\
        SN 2019muj & -16.40 &  \citet{Barna21a} & 4600 & \citet{Barna21a}\\
        SN 2022xlp & -16.06 &  \citet{Banhidi25} & 4700 & \citet{Banhidi25}\\          
        SN 2023zgx & -14.70 &  \citep{Barna26} & 3100 & \citep{Barna26} \\
        SN 2020kyg & -14.50 &  \citet{Singh23} & 3000 & \citet{Singh23}\\ 
        SN 2019gsc & -13.97 &  \citet{Srivastav20} & 2100 & \citet{Srivastav20} \\
        SN 2022ywf & -13.81 &  \citet{Barna26} & 2400 & \citet{Barna26} \\
            \hline
\end{longtable}
\end{center}

    \begin{table}[!h]
      \caption{The settings and best-matching parameters of the TARDIS model of SN 2024pxl } 
      \centering
        \label{tab:TARDIS-settings}
        \begin{tabular}{l | l } 
            Physical treatments     &       \\
            \hline
            \noalign{\smallskip}
            ionization mode         &       nebular         \\
            excitation mode         &       dilute-lte      \\
            radiative rates type    &       dilute-blackbody\\
            line interaction type   &       macroatom       \\
            \noalign{\smallskip}
            \hline
            \noalign{\smallskip}
            Number of photon packets    &   50000           \\
            Number of final pakcets     &   150000          \\
            Iterations                  &   15              \\
            \noalign{\smallskip}
            \hline
            \noalign{\smallskip}
            Density function        &       \\
            \hline
            \noalign{\smallskip}
            $\rho_\mathrm{0}$ [g\,cm$^{-3}$]      &    0.2  \\
            $v_\mathrm{0}$ [km\,s$^{-1}$]        &    3000               \\
            $v_\mathrm{cut}$ [km\,s$^{-1}$]        &    6500               \\
            $C$                                 &    $8.2 \times 10^{5}$ \\
            Outer velocity boundary         &   $v_\mathrm{phot} + 5000$   \\
            $T_\mathrm{exp}$ [MJD]              &    $60510.6$ \\
            \noalign{\smallskip}
            \hline
            
         \end{tabular} 
    \end{table}

\end{document}